\documentclass[10pt,twocolumn,letterpaper]{article}
\usepackage{cvpr} 

\usepackage{microtype}
\usepackage{multirow}
\usepackage{multicol}
\usepackage{xcolor}
\usepackage{amssymb}
\usepackage{pifont}
\usepackage{caption}
\usepackage{graphicx}
\usepackage{array}
\usepackage{makecell}
\usepackage{booktabs}
\usepackage{siunitx}
\usepackage[accsupp]{axessibility}

\newcommand{\xmark}{\ding{55}}

\definecolor{cvprblue}{rgb}{0.21,0.49,0.74}
\usepackage[pagebackref,breaklinks,colorlinks,allcolors=cvprblue]{hyperref}

\def\paperID{70}
\def\confName{CVPR}
\def\confYear{2026}

\title{OGA-AID: Clinician-in-the-loop \underline{AI} Report \underline{D}rafting Assistant \\for Multimodal \underline{O}bservational \underline{G}ait \underline{A}nalysis in Post-Stroke Rehabilitation}

\author{
Khoi T. N. Nguyen$^{1, 2}$ \and
Nghia D. Nguyen$^{3,5}$ \and
Koh Hui Yu$^{1, 2}$ \and
Patrick W.\ H.\ Kwong$^{4}$ \and
Karen Sui Geok Chua$^{1,6}$ \and
Ananda Sidarta$^{1,2*\dagger}$ \and
Baosheng Yu$^{2}$\thanks{Senior authors. $\dagger$ Corresponding author: ananda.sidarta@ntu.edu.sg} \and
\\ \small $^{1}$Rehabilitation Research Institute of Singapore, Nanyang Technological University, Singapore \\ \small $^{2}$Lee Kong Chian School of Medicine, Nanyang Technological University, Singapore \\ \small
$^3$The Grainger College of Engineering, University of Illinois Urbana-Champaign, United States \\ \small
 $^4$ Department of Rehabilitation Sciences, The Hong Kong Polytechnic University, Hong Kong
\\ \small $^5$VinUni-Illinois Smart Health Center, VinUniversity, Vietnam \\ \small $^6$Institute of Rehabilitation Excellence, Tan Tock Seng Hospital, NHG Health, Singapore
}

\begin{document}
\maketitle
\begin{abstract}
Gait analysis is essential in post-stroke rehabilitation but remains time-intensive and cognitively demanding, especially when clinicians must integrate gait videos and motion-capture data into structured reports. We present OGA-AID, a clinician-in-the-loop multi-agent large language model system for multimodal report drafting. The system coordinates 3 specialized agents to synthesize patient movement recordings, kinematic trajectories, and clinical profiles into structured assessments. Evaluated with expert physiotherapists on real patient data, OGA-AID consistently outperforms single-pass multimodal baselines with low error. In clinician-in-the-loop settings, brief expert preliminary notes further reduce error compared to reference assessments. Our findings demonstrate the feasibility of multimodal agentic systems for structured clinical gait assessment and highlight the complementary relationship between AI-assisted analysis and human clinical judgment in rehabilitation workflows.
\end{abstract}
    
\vspace{-25pt}
\section{Introduction}
\label{sec:intro}
Stroke often causes long-term movement difficulties, and rehabilitation requires ongoing assessment to understand how a patient is recovering. In physiotherapy practice, \emph{observational gait analysis} (OGA), in which physiotherapists observe patients' gait patterns from their movements, is commonly performed to identify abnormalities and functional improvements to inform intervention planning. To support a more objective assessment beyond visual observation alone, motion capture systems (MoCap) and wearable sensors have been increasingly adopted~\cite{Unravelingstrokegait, van_criekinge_full-body_motion_capture_2023, SHARMA2024228} to quantify kinematic and spatiotemporal gait characteristics, such as joint angles, step symmetry, and temporal coordination. While these technologies provide detailed and precise measurements, their outputs are often time-consuming and require extensive effort to interpret, especially without biomechanics expertise~\cite{SHARMA2024228}. Consequently, rather than reducing workload, such systems may increase the cognitive burden for physiotherapists who must manually translate complex multidimensional data into clinically meaningful reports.

Recent advances in large language models (LLMs) and vision language models (VLMs) have led to the rapid adoption of AI assistants across a wide range of medical domains, including clinical documentation, medical question answering, and decision support~\cite{Ren2025HealthcareAgent, vrdoljak2025review}. These models have demonstrated strong capabilities in summarizing complex information, organizing heterogeneous inputs, and generating structured textual outputs. However, the application of LLM-based systems to OGA, especially for post-stroke rehabilitation, remains largely unexplored. To the best of our knowledge, no previous study has systematically examined the role of LLM-based systems in this clinical assessment setting. This gap is particularly salient given physiotherapists' interest in AI-assisted workflows~\cite{Alwhaibi2025} that enhance efficiency while retaining human oversight to ensure outputs can be validated before being shared with patients.

Motivated by these challenges, we introduce \textbf{OGA-AID}, an AI report assistant for post-stroke OGA. OGA-AID integrates the patient's movement recordings, MoCap joint trajectories, clinical profile, and clinician's initial observations to produce structured gait assessment draft reports articulated in clear clinical language. Crucially, OGA-AID is designed to be an assistive tool rather than a replacement for clinical judgment, in which clinicians retain full control by reviewing and finalizing the generated reports. 

OGA-AID employs an agentic AI design consisting of 3 specialized agents with distinct roles: a \textbf{Recording Observer} that analyzes temporal movement recordings from frontal and sagittal (side) viewpoints; a \textbf{Trajectory Analyzer} that processes and interprets MoCap data; and a \textbf{Report Generator} that synthesizes these insights into a formatted clinical report. The resulting AI-generated reports are delivered through a user-friendly interface that allows clinicians to inspect, modify, and control the final outputs.

We evaluate OGA-AID through a clinician-centered study with a partner hospital using real-world post-stroke patient data. Our evaluation covers both autonomous and clinician-in-the-loop performance with deep involvement of expert physiotherapists. In summary, our contributions are:
\begin{itemize}
\item We introduce \textbf{OGA-AID}, a clinician-in-the-loop AI report assistant for post-stroke OGA. The system integrates different input modalities, including patient movement videos, MoCap trajectories, and clinical profiles. We further propose a structured decomposition of different gait factors and assign them to specialized agents to effectively analyze post-stroke gait.

\item We conduct clinician-centered evaluations using real-world post-stroke patient data with diverse demographics and post-stroke durations. Results show that generated reports achieve clinically acceptable alignment with expert physiotherapists, paving the way for assistive agentic AI in post-stroke rehabilitation practice.

\item Through our experiments, we show that clinician involvement in the OGA-AID pipeline, such as providing preliminary observations, improves the quality of the generated reports. This finding highlights the importance of human-AI collaboration and supports clinician-in-the-loop designs for AI-assisted analysis workflows.
\end{itemize}

\section{Background \& Related Work}
\label{sec: related_works}

\textbf{Post-stroke OGA and the Wisconsin Gait Scale.} In post-stroke rehabilitation, OGA is essential for identifying movement asymmetry and abnormalities such as hip circumduction and knee hyperextension, which are critical when tailoring interventions. These observations are critical for comprehensive assessments, treatment planning, and monitoring recovery progress~\cite{10.2522/ptj.20120344}. To mitigate subjectivity and improve consistency, several standardized clinical gait scales have been developed. Among these, the Wisconsin Gait Scale (WGS)~\cite{Rodriquez1996Gait} is designed to assess hemiplegic stroke gait. The WGS comprises 14 observable items that span both the stance and swing phases, each accompanied by clear descriptive criteria to guide assessment. For each factor, observed gait deviations are translated into ratings, mostly from 1 (normal) to 3 (atypical), based on predefined severity levels. The total WGS score, which is the weighted sum of all factor ratings, ranges from 13.35 to 42, with higher scores reflecting greater gait impairments.

\textbf{MoCap-Assisted Gait Analysis.} To capture movement traits that are difficult to observe with human eyes, MoCap is adopted for quantitative gait analysis~\cite{Unravelingstrokegait, van_criekinge_full-body_motion_capture_2023, SHARMA2024228}. MoCap systems are commonly categorized as marker-based or markerless. Marker-based systems reconstruct 3D motion using reflective markers placed on anatomical landmarks and are widely regarded as the gold standard for biomechanical analysis due to their accuracy. In contrast, markerless systems improve accessibility by leveraging computer vision and deep learning, eliminating the need for physical markers and reducing setup cost. However, their accuracy can be affected by occlusion, lighting conditions, or atypical movement patterns. Depending on available resources and infrastructure, clinical centers may adopt either approach for gait analysis, acknowledging the inherent trade-offs. 

As OGA relies on visual judgment, its measurements can vary substantially across clinicians~\cite{matsuzaka2025reliability}. Using MoCap, certain gait factors, such as movements of the hip, knee, ankle, and trunk, can be measured more reliably. In MoCap-assisted post-stroke gait assessment, a common practice is to compare a patient's gait with normative gait patterns from healthy individuals~\cite{diagnostics15050511} matched by demographics such as age, height, weight, and gender. This comparison allows clinicians to contextualize observed impairments and identify functional deviations across different body segments.

\textbf{LLMs and Agentic AI for Healthcare.} Recent advances in LLMs and VLMs have accelerated the development of AI assistants across a wide range of healthcare applications. State-of-the-art foundation models, such as GPT~\cite{singh2025openaigpt5card} and Gemini~\cite{comanici2025gemini25pushingfrontier}, are trained on large-scale multimodal data including medically relevant information, enabling strong capabilities in medical question answering, clinical documentation, and decision support. Building on these capabilities, recent studies have explored agentic paradigms that coordinate multiple specialized models to support healthcare workflows~\cite{liu2025medmmvcontrollablemultimodalmultiagent, Ren2025HealthcareAgent, vrdoljak2025review}. Within stroke rehabilitation, emerging work has begun to investigate AI systems for clinical education, therapy assistance, and clinician–patient interaction~\cite{nguyen2025congaitcliniciancentereddashboardcontestable, ethstrokerehabagent, Qiang2025}. However, these systems are primarily designed around conversational interfaces that provide guidance or educational feedback rather than gait analysis. They remain largely text-driven and make limited use of the patient's gait or motion data. As a result, the potential of LLM-based systems to directly support clinicians in OGA remains largely unexplored.

\begin{figure*}[ht]
  \centering
  \includegraphics[width=\linewidth]{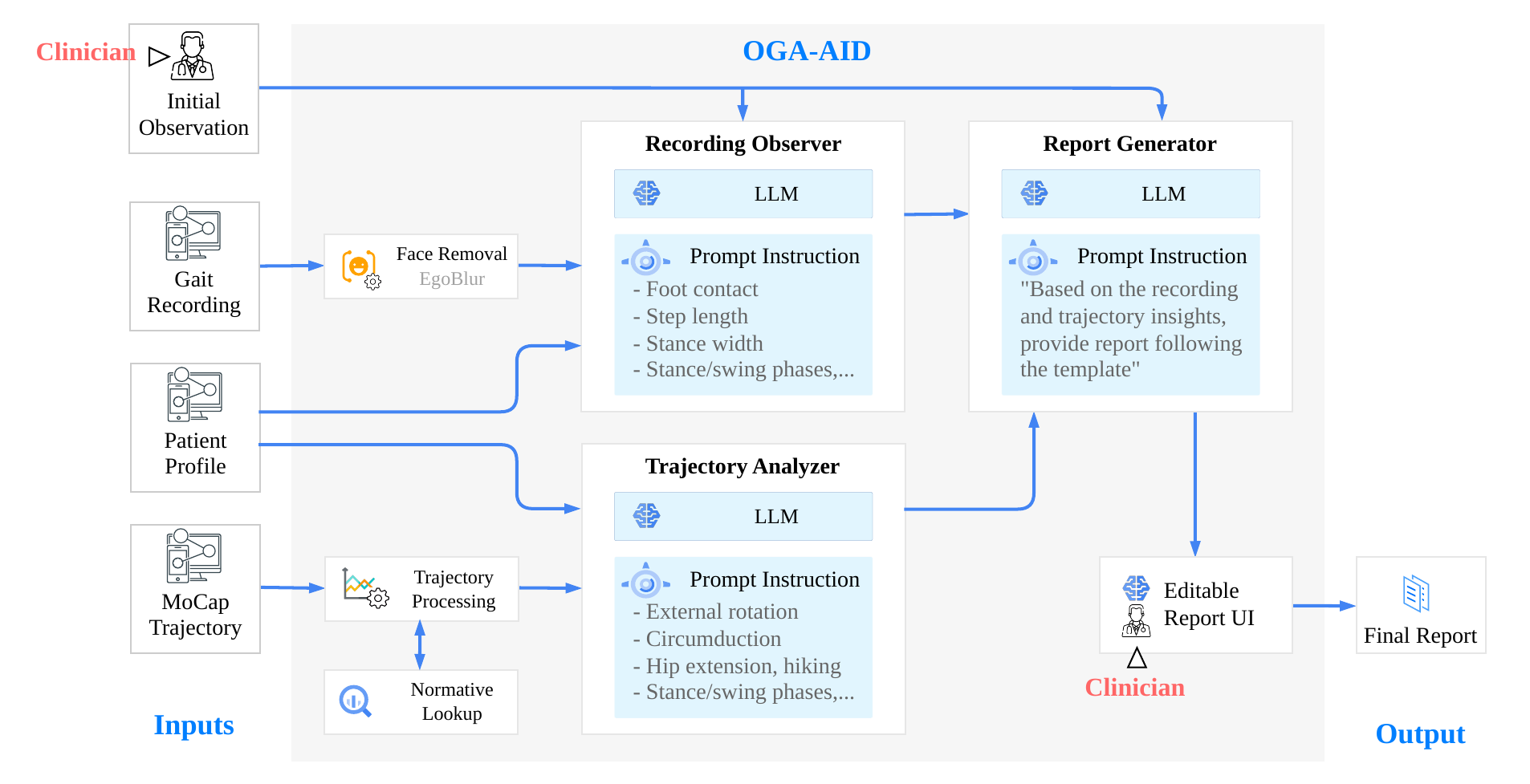} 
  \vspace{-20pt}
  \caption{The overall architecture of OGA-AID. The system is designed to integrate multi-modal gait evidence with contextual patient information and clinician preliminary observations for post-stroke gait assessment through coordinated, agent-based analysis.}
  \label{fig:pipeline}
  \vspace{-10pt}
\end{figure*}

\section{Methodology}
\label{sec:approach}

\subsection{Overview}

Figure~\ref{fig:pipeline} illustrates OGA-AID's architecture. To ensure clinical interpretability, our agentic pipeline integrates WGS to guide gait analysis and standardize the output template. Note that, instead of WGS, other OGA instruments can also be integrated. For gait analysis, individual WGS factors are assigned to the most appropriate agent instead of jointly processing all factors at once. Similar to standard practice mentioned in Section~\ref{sec: related_works}, OGA-AID matches patients' gait patterns against normal individuals by retrieving from a database of gait trajectories collected within the hospital. This information is processed within OGA-AID for downstream LLM-based report generation.

Importantly, unlike fully automated AI-based approaches, OGA-AID is designed to support clinician-in-the-loop report generation grounded in multimodal gait evidence. In addition to the 3 default inputs - gait recordings, MoCap trajectories, and patient clinical profile - clinicians may provide preliminary notes based on their initial observations. These lightweight observations are incorporated as context references for the agents. After the report is generated, clinicians review and finalize the content prior to patient communication. In clinical environments where maintaining professional authority remains essential, we believe this collaborative design offers a balanced and safe approach to integrating AI into practice.

\subsection{Input Details}
\label{ssec: input_acquisition}

In a typical MoCap-assisted OGA, the patient performs a 10-meter walking task at a comfortable pace along a clear walkway. To capture motion trajectories, reflective markers are attached to the patient during the task. We assume that the clinical environment is equipped with a functional MoCap system. OGA-AID operates with 4 inputs as follows:

\textbf{Gait recording.} During the walking task, patient movements are typically captured from multiple camera views. For OGA-AID, we utilize recordings from two views, frontal and sagittal (side), which are sufficient for visual gait assessment while reducing video token usage.

\textbf{MoCap trajectory.} Markerless or marker-based trajectories can be obtained from the patient's gait movements using third-party software. The raw MoCap data comprise time-series 3D marker coordinates, which are internally calibrated within the software. These trajectories are provided to OGA-AID as input for downstream analysis.

\textbf{Patient clinical profile.} This input comprises the patient’s clinical characteristics, including demographic attributes (age, sex, height, and weight) as well as the affected hemiparetic side. Identification of the hemiparetic side is particularly critical, as it directly influences accurate gait interpretation in standard gait analysis.

\textbf{Clinician's initial observation.} In practice, a collaborative agentic system like OGA-AID rarely works independently without inputs from the clinician. Clinicians may optionally provide brief free-text notes for additional context, reflecting their initial impressions (e.g., ``no gait aid", ``prolonged stance phase"). In particular, some gait factors can be straightforwardly observed from a patient recording without creating a cognitive burden. These observations can serve as supplementary context for the agents and reduce the overall task complexity, enabling the agent to focus on the remaining items with greater efficiency.

\begin{figure}[ht]
  \centerline{\includegraphics[width=\linewidth]{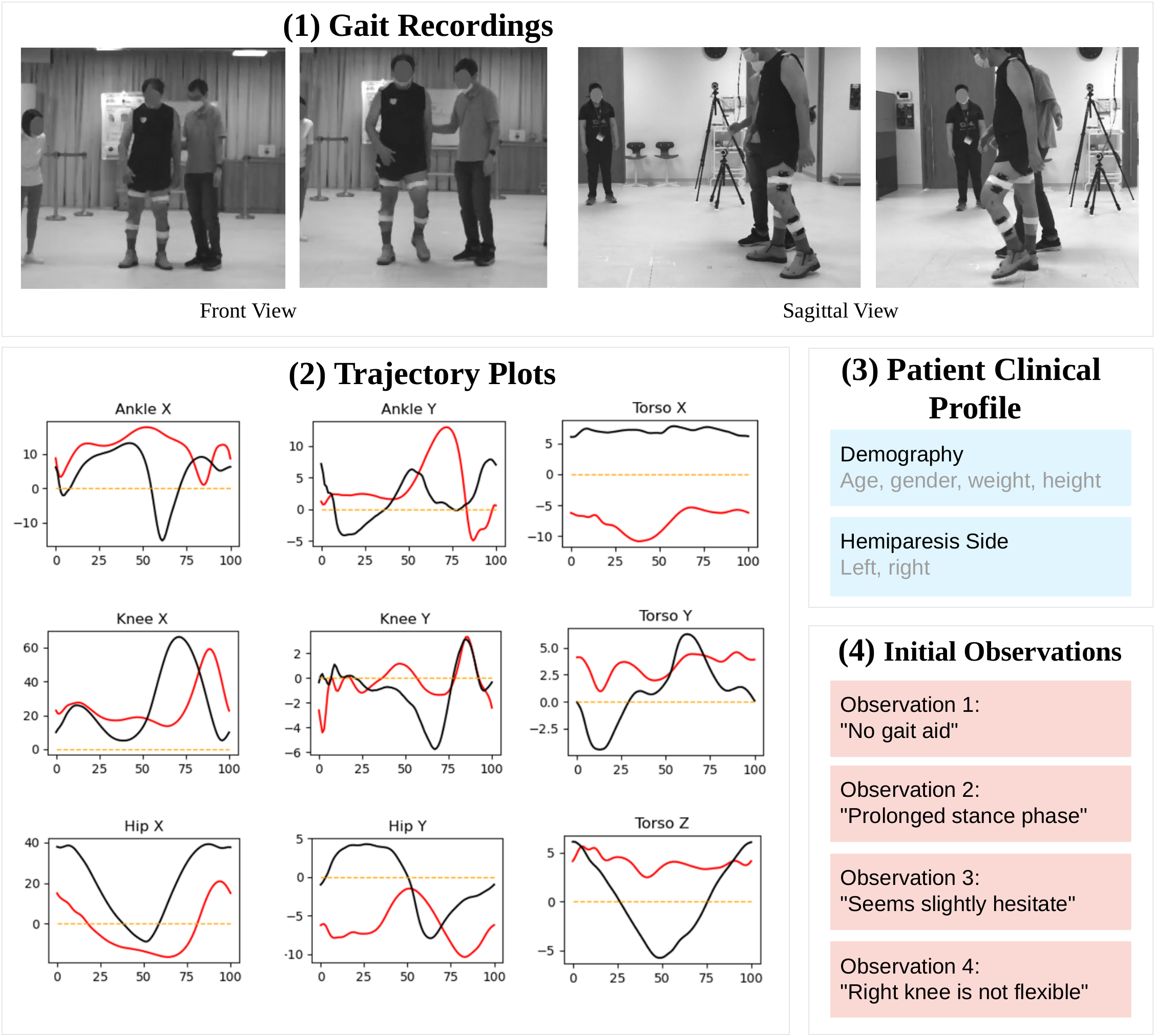}}
  \caption{Processed multi-modal inputs for LLM inference in OGA-AID. (1) Anonymized gait recording that blurred the faces of people, (2) Joint trajectory plots after processing the MoCap trajectory input, (3) Patient clinical profile with demography and hemiparesis information, (4) Clinician's initial observation.}
  \vspace{-5pt}
  \label{fig:inputs_fig}
  \vspace{-10pt}
\end{figure}

\subsection{Input Processing}

In OGA-AID, the input data is processed before being sent to the LLM for analysis, as shown in Figure~\ref{fig:inputs_fig}.

\textbf{Face removal.} To meet Institutional Review Board standards, we apply face removal before sending data to the LLMs. We employ EgoBlur~\cite{raina2023egoblur}, which automatically detects and blurs the faces of all individuals in the video, including the stroke patient and assisting personnel. This processing step reduces the risk of identity disclosure while preserving gait-relevant visual information for analysis.

\textbf{Normative lookup.} The patient profile is used to retrieve corresponding normative references from an established database, assuming such reference gait data are available. Matching is performed using basic demographics, i.e., age, sex, height, and weight. In our setting, we are provided with gait trajectories from 50 normative subjects in advance.

\textbf{Trajectory processing.} The raw MoCap data consist of time-series sequences of 3D marker coordinates. The trajectories of the stroke patient and the matched normative subject are processed as spatiotemporal plots for the hip, knee, ankle, and torso joints. To enable consistent comparison across walking trials, the processed joint trajectories are time-normalized to a 0-100\% gait cycle, allowing the agent to localize deviations within specific gait phases. For example, a downward deviation in the ankle trajectory during the mid-swing phase (61-80\%) may suggest inadequate toe clearance. These processed trajectories are then visualized as spatiotemporal plots, highlighting potential gait abnormalities such as reduced joint range of motion or irregular temporal patterns. In Figure~\ref{fig:inputs_fig}, the red curves represent the patient trajectories, while the black curves correspond to the matched normative subject. Such visualizations provide interpretable motion context that supports subsequent observational analysis and report generation.

\subsection{Multi-Agent Architecture and Agent Roles}
\begin{table*}[t]
\centering
\caption{Assessment factors for post-stroke gait defined in the WGS. Each factor is evaluated using an ordinal scale reflecting increasing severity. OGA-AID utilizes these factors to generate structured gait analysis reports.}
\label{tab:wgs_factors}

\begin{minipage}[t]{0.525\textwidth}
\vspace{-10pt}
\centering
\footnotesize
\renewcommand{\arraystretch}{1.2}
\setlength{\tabcolsep}{1pt}
\begin{tabular}{@{}>{\raggedright\arraybackslash}p{0.20\textwidth}p{0.80\textwidth}@{}}
\multicolumn{2}{c}{\textbf{Recording-based Factors}}\\
\hline
\textbf{Factor} & \textbf{Description} \\
\hline

Use of \linebreak hand gait aid & Dependence on assistance during walking reflects impaired balance, weight transfer, or lower limb control. \\

Unaffected side \linebreak step length & Shortened forward step distance reflects inadequate weight transfer, weakness, or instability of the affected limb. \\

Affected side weight shift & Insufficient body-weight shift toward paretic limb suggests impaired balance or weakness in the anti-gravity muscles of the lower limb. \\

Stance width & Mediolateral distance between feet during walking. Widened distance suggests balance deficits. \\

Guardedness & Hesitant gestures like stiffness, reduced arm swing, or slow movement due to fear of falling or difficult movement coordination. \\

Initial swing \linebreak knee flexion & Reduced flexion typically leads to poor foot clearance and compensatory strategies. \\

Toe clearance & If toes adequately clear the ground during swing. Inadequate clearance increases tripping risks. \\

Initial \linebreak foot contact & How the affected foot contacts the ground. Abnormal patterns reflect impaired ankle control. \\

\hline
\end{tabular}
\end{minipage}
\hfill
\begin{minipage}[t]{0.45\textwidth}
\vspace{-10pt}
\centering
\footnotesize
\renewcommand{\arraystretch}{1.35}
\setlength{\tabcolsep}{1pt}
\begin{tabular}{@{}>{\raggedright\arraybackslash}p{0.25\textwidth}p{0.73\textwidth}@{}}
\multicolumn{2}{c}{\textbf{Trajectory-based Factors}}\\
\hline
\textbf{Factor} & \textbf{Description} \\
\hline

Affected side \linebreak stance time & Reduced stance duration indicates instability in loading the affected side. \\

Affected side \linebreak hip extension & Hip movement during terminal stance. Limited extension reduces propulsion and disrupts gait rhythm. \\

Affected leg external rotation & Excessive outward rotation of the paretic hip during swing, potentially arising from weakness or impaired motor control. \\

Mid-swing circumduction & If present, affected limb swings outward in a semi-circular motion to compensate for reduced hip flexion, knee flexion, or ankle dorsiflexion.  \\

Mid-swing \linebreak hip hiking & Elevation of the pelvis on the affected side during swing, typically to compensate for reduced hip flexion, knee flexion, or ankle dorsiflexion.  \\

Terminal swing \linebreak pelvic rotation &  Limited rotation suggests trunk stiffness or impairments in lower limb strength or motor control.\\

\hline
\end{tabular}
\end{minipage}
\vspace{-10pt}
\end{table*}

In OGA, physiotherapists assess multiple gait-related factors before producing a clinical report. Using the WGS as the reference standard, we incorporate 14 assessment factors. Although the factors can be examined visually, trajectory data offer a more objective and quantitative characterization of gait kinematics. In consultation with clinical experts, we use video recordings to analyze 8 visually dominant factors, whereas the remaining 6 factors are assessed using MoCap trajectories for more precise quantitative measurement. To systematically handle diverse factors and heterogeneous inputs, we adopt a multi-agent architecture with a factor decomposition strategy. Table~\ref{tab:wgs_factors} details the description and allocation of WGS factors to each agent. To ensure consistency across agents, the patient's clinical profile, including age, gender, and the hemiparetic side, is provided to all agents. Details of the agents are as follows:

\textbf{Recording Observer.} This agent analyzes gait recordings captured from frontal and sagittal views. To process the video data within the model’s context limits, we employ an equidistant frame-sampling strategy, extracting 50 frames uniformly across the 10-meter walking sequence. Since clinical recordings may include assisting personnel (e.g., therapists), the agent is prompted to perform zero-shot subject identification using the patient’s clinical profile. For example, the agent may identify the stroke patient by matching the described hemiparetic side with observable gait asymmetry. After identifying the patient, the agent performs temporal reasoning across the sampled frames to generate observations for 8 visually assessed factors, analyzing the spatial relationships between the patient’s limbs and the surrounding environment. When MoCap trajectory input is unavailable, the system switches to a video-only pipeline, in which the Recording Observer evaluates all 14 WGS factors directly from the gait recordings.

\textbf{Trajectory Analyzer.} This agent analyzes 3D gait kinematics by interpreting MoCap joint trajectory plots of the hip, knee, ankle, and torso along 3 orthogonal planes. To standardize the input for the LLM, the agent interprets trajectory displacement along the normalized gait cycle. The agent is prompted to perform a comparative analysis between the patient’s impaired-side trajectory (red curve) and the matched normative baseline (black curve). By examining differences in curve magnitude, peak locations, and slope transitions across the stance phase and swing phase, the agent identifies characteristic stroke-related motion patterns, such as reduced hip extension, altered stance duration, or pelvic retraction. These detected deviations are then translated into qualitative categorical descriptors, producing structured assessments for the remaining 6 WGS factors.

\textbf{Report Generator.} This agent aggregates the outputs from the Recording Observer and Trajectory Analyzer and reconciles with any clinician's initial observations to produce a formatted clinical draft that follows either a factor-specific WGS scoring template or a narrative clinical summary. This design ensures the generated report remains both clinically interpretable and easily editable by clinicians.

\subsection{LLM Selection and Prompt Strategy}

We select multimodal LLMs with strong image and video understanding capabilities, which in fact only exists a few suitable candidates. In Section~\ref{sec: evaluation}, we study 4 models: Qwen-3 VL (\texttt{qwen3-vl-plus})~\cite{bai2025qwen3vltechnicalreport}, GPT-5~\cite{singh2025openaigpt5card} including GPT-5 Mini (\texttt{gpt-5-mini}) and GPT-5.1 (\texttt{gpt-5.1}), and Gemini 3 Flash (\texttt{gemini-3-flash-preview})~\cite{NewEraIntelligence2025}. We intentionally exclude extended-reasoning or thinking variants, as the primary objective of OGA-AID is structured clinical analysis and report generation rather than exploratory reasoning. Instruction-following models are more suitable for producing consistent, template-aligned outputs~\cite{li2025when} grounded in provided evidence. For consistency and fair evaluation of the LLM capability, all agents within OGA-AID employ the same base LLM. Nevertheless, future work may optimize LLM combinations or integrate newer LLMs for further improvements.

Prompting is standardized using structured, role-based instructions aligned with the WGS. When both MoCap trajectories and gait recordings are available, the Recording Observer operates under a constrained assessment prompt that focuses exclusively on the 8 visually observable WGS factors. By restricting the prompt scope to the relevant visual attributes, this configuration improves the reliability of qualitative observations. If the MoCap trajectory is unavailable, the Recording Observer evaluates all 14 WGS factors directly from recordings using a full assessment prompt.
\section{Experiment Design}
\label{sec: experiments}

\subsection{Expert Participation}

Two experienced physiotherapists were involved in both the experimental procedure and evaluation.

\begin{figure*}[ht]
  \centering
  \includegraphics[width=\linewidth]{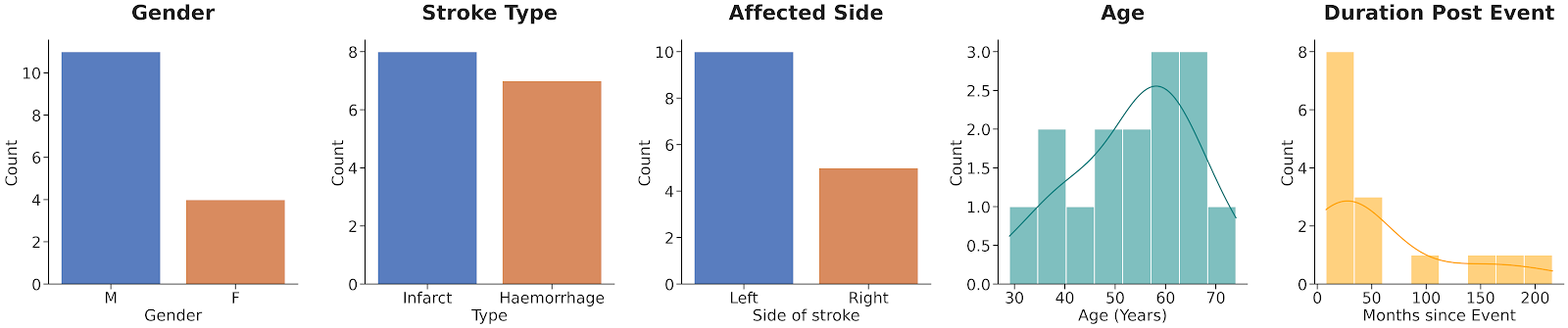} 
  \vspace{-15pt}
  \caption{Statistics of 15 patients, including gender, stroke type, hemiparetic side, age, and duration since the stroke event (in months).}
  \label{fig:patient_clinical_profile}
  \vspace{-10pt}
\end{figure*}

\textbf{\includegraphics[height=1.75ex]{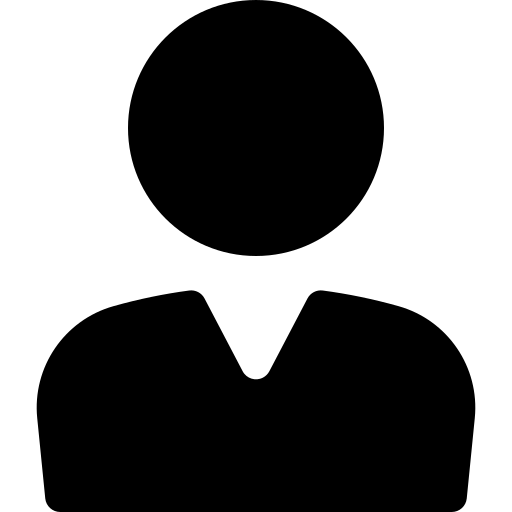}~Physiotherapist A}: Senior physiotherapist with 8 years of experience in rehabilitation in both community and private sectors, working with adults presenting neurological, orthopaedic, and musculoskeletal conditions.

\textbf{\includegraphics[height=1.75ex]{figs/human_icon.png}~Physiotherapist B}: Principal physiotherapist with 15 years of experience in community rehabilitation and vocational reintegration for people with physical disabilities. 

Additionally, a senior biomechanist was involved to support MoCap interpretation and facilitate the provision of the reference WGS made by the physiotherapists.

\subsection{Experiment Details}

Our experiments involved 15 post-stroke survivors, with statistics demonstrated in Figure~\ref{fig:patient_clinical_profile}. The cohort includes a balanced distribution of stroke types, with ages ranging from 30 to 80 years and varying chronic stages from 8 to 216 months post-event. To ensure statistical robustness despite the clinical constraints of recruiting post-stroke participants, each participant is asked to complete the walking task 3 times. This yielded a total of $N=45$ data samples, capturing intra-patient variability in movement execution and environmental noise (e.g., differing positions and numbers of assisting personnel). The input acquisition follows the procedure described in Section~\ref{ssec: input_acquisition}. In our setting, we employed a 16-camera Qualisys Miqus-M3~\cite{QualisysQTM}, an industry-leading marker-based MoCap software, to collect precise kinematic trajectories. However, alternative MoCap systems can be used depending on budget conditions.

\textbf{Reference WGS.} By analyzing the gait recordings and the trajectory data, participating experts work together to conduct the WGS reports. To ensure a more objective measurement, both physiotherapists must obtain mutual agreement for all 15 patients. We consider the WGS reports by the physiotherapists as a reference to assess OGA-AID.

\textbf{LLM Configurations and Multi-run Inferences.} We set the temperature to 0.0 for all LLMs to encourage deterministic behavior, except for GPT-5 Mini, which does not support temperature control and operates with a fixed default setting. Nevertheless, multimodal LLMs may still exhibit nondeterministic behavior due to internal stochastic processes and vision processing mechanisms. Slight variations may arise from token selection dynamics or the internal handling of video or image sequences. To reduce the impact of such variability, for each of $N=45$ samples, we conduct $M=3$ independent runs and average the results.

\subsection{Autonomous Evaluation}
\label{ssec: auto_metric}

Section~\ref{ssec: wgs_eval} examines the performance of OGA-AID in generating WGS reports without clinician-in-the-loop, enabling an independent assessment of its clinical reasoning and structured report generation. We compare OGA-AID with single-pass LLM baselines, which receive all 3 input modalities (gait recordings, MoCap trajectories, and the patient's clinical profile) within a single prompt. Using the same instruction prompt template as OGA-AID, the model directly outputs assessments for all 14 WGS factors and produces the report in a single inference step. We further investigate the effect of different input combinations in OGA-AID to better understand how each input modality contributes to the accuracy of the generated reports.

\subsection{Clinician-in-the-loop Evaluation}
\label{ssec: cil_metric}

Section~\ref{ssec: cil_eval} evaluates OGA-AID in a collaborative setting where clinicians provide preliminary observations to guide the analysis. To simulate this scenario, each patient is associated with 5 observations that physiotherapists can straightforwardly identify, including knee flexion during initial swing, circumduction, stance time, hip extension, and step length. Each run uses a different sampling of these observations. The sampled observations are categorized into \textit{short} (1–2 observations), \textit{medium} (3–4 observations), and \textit{long} (5 observations). This evaluation measures the performance gains enabled by the clinician-in-the-loop setting.

\subsection{Metrics}
\label{ssec: metrics}

Let $y_i$ denote the reference WGS score and $\hat{y}_i$ denote the predicted score for patient $i$. We report 4 metrics as follows.

\textbf{Mean Absolute Error (MAE)} quantifies the average magnitude of prediction error across all samples, defined by
$\frac{1}{NM} \sum_{n=1}^{N} \sum_{m=1}^{M} \left| \hat{y}_{n,m} - y_{n} \right|.$ We report the MAE averaged over 45 data samples and 3 runs per sample. Lower MAE indicates closer agreement with the reference WGS. 

\textbf{Maximum Absolute Error (Max AE)} captures the largest deviation, defined by $\max_{n,m} \left| \hat{y}_{n,m} - y_n \right|.$ We report the largest Max AE across $N \times M=135$ runs to reflect the most extreme potential error. This metric reflects worst-case discrepancies, which are clinically important, as extreme deviations may affect treatment decisions. 

\textbf{Mean Error (Bias)} measures systematic over- or under-estimation (positive or negative value), defined by $\frac{1}{NM} \sum_{n=1}^{N} \sum_{m=1}^{M} \left( \hat{y}_{n,m} - y_{n} \right).$ We report the final Bias averaged over the 45 data samples and 3 runs per sample. 

\textbf{Minimal Clinically Important Difference (MCID)} represents the smallest change in a score that is perceived as \textit{clinically meaningful} in the patient's status. In our case, we are interested in understanding whether the conclusions made by the agentic system can be assistive to clinician workflow, i.e, do not result in large disparity compared to physiotherapists' evaluations. Note that in OGA-AID, clinicians preserve control of the final outputs, so perfect agreement is not necessary, and agreement metrics are less informative. Therefore, we instead evaluate whether the difference reported is less than a threshold defined by the MCID. Differences exceeding this threshold are considered inappropriate, as they may require the clinician to fully reevaluate the scores, which significantly impacts the usability of the system. For WGS, the MCID is defined as 2.25~\cite{guzikWisconsinGaitScale2019}.

\begin{table}[htbp]
\centering
\caption{Performance comparison between single-pass LLMs and OGA-AID for WGS score reports for $N=45$ samples, without clinician-in-the-loop. $\downarrow$ indicates lower values are better. ``Below MCID" denotes if the MAE falls under the 2.25 MCID threshold.}
\vspace{-5pt}
\label{tab:main_table}
\setlength{\tabcolsep}{3pt}
\begin{tabular}{llcccc}
\toprule
& LLM 
& MAE $\downarrow$ 
& Max AE $\downarrow$ 
& Bias 
& \makecell{Below \\ MCID} \\
\midrule

\multirow{4}{*}{\makecell{Single\\pass}}
& Qwen-3 VL      & 9.18 & 20.00 & 9.18 & \textcolor{BrickRed}{\xmark}\\
& GPT-5 Mini     & 3.15 & 7.70 & 1.61 & \textcolor{BrickRed}{\xmark} \\
& GPT-5.1        & 2.35 & 4.75 & 0.45 & \textcolor{BrickRed}{\xmark}\\
& Gemini-3 Flash & 2.41 & 5.58 & 0.95 & \textcolor{BrickRed}{\xmark} \\

\midrule

\multirow{4}{*}{\begin{tabular}{c} Ours \end{tabular}}
& Qwen-3 VL      
    & 6.28 & 11.00 & 6.28 
    & \textcolor{BrickRed}{\xmark}\\
& GPT-5 Mini
    & 3.17 & 8.00 & 3.04
    & \textcolor{BrickRed}{\xmark} \\
& GPT-5.1        
    & 2.05 & 6.75 & 0.38  
    & \textcolor{ForestGreen}{$\checkmark$}\\
& Gemini-3 Flash 
    & 1.94 & 4.25 & 0.23  
    & \textcolor{ForestGreen}{$\checkmark$}\\

\bottomrule
\end{tabular}
\vspace{-5pt}
\end{table}

\section{Results}
\label{sec: evaluation}

\subsection{Autonomous Evaluation}
\label{ssec: wgs_eval}

Table~\ref{tab:main_table} presents a comparison between OGA-AID and single-pass LLM baselines. Across all backbones, OGA-AID generally achieves substantially lower MAE than single-pass inference, highlighting the advantage of the structured, multi-stage generation pipeline for the OGA task. Notably, with GPT-5.1 and Gemini-3 Flash, OGA-AID attains MAE values below the 2.25 MCID threshold, indicating clinically acceptable agreement with reference scoring. We find these results encouraging, as they suggest that \textbf{agentic AI systems and LLM-based frameworks, such as OGA-AID, may offer a promising direction for supporting post-stroke gait analysis}. More broadly, we highlight the potential role of LLM-driven tools in rehabilitation assessment workflows, an important yet relatively underexplored domain that requires greater research attention. Nonetheless, the relatively significant Max AE also provides a warning about cautious use of agentic systems since they may provide misaligned predictions.

\begin{figure}[ht]
  \centerline{\includegraphics[width=1.1\linewidth]{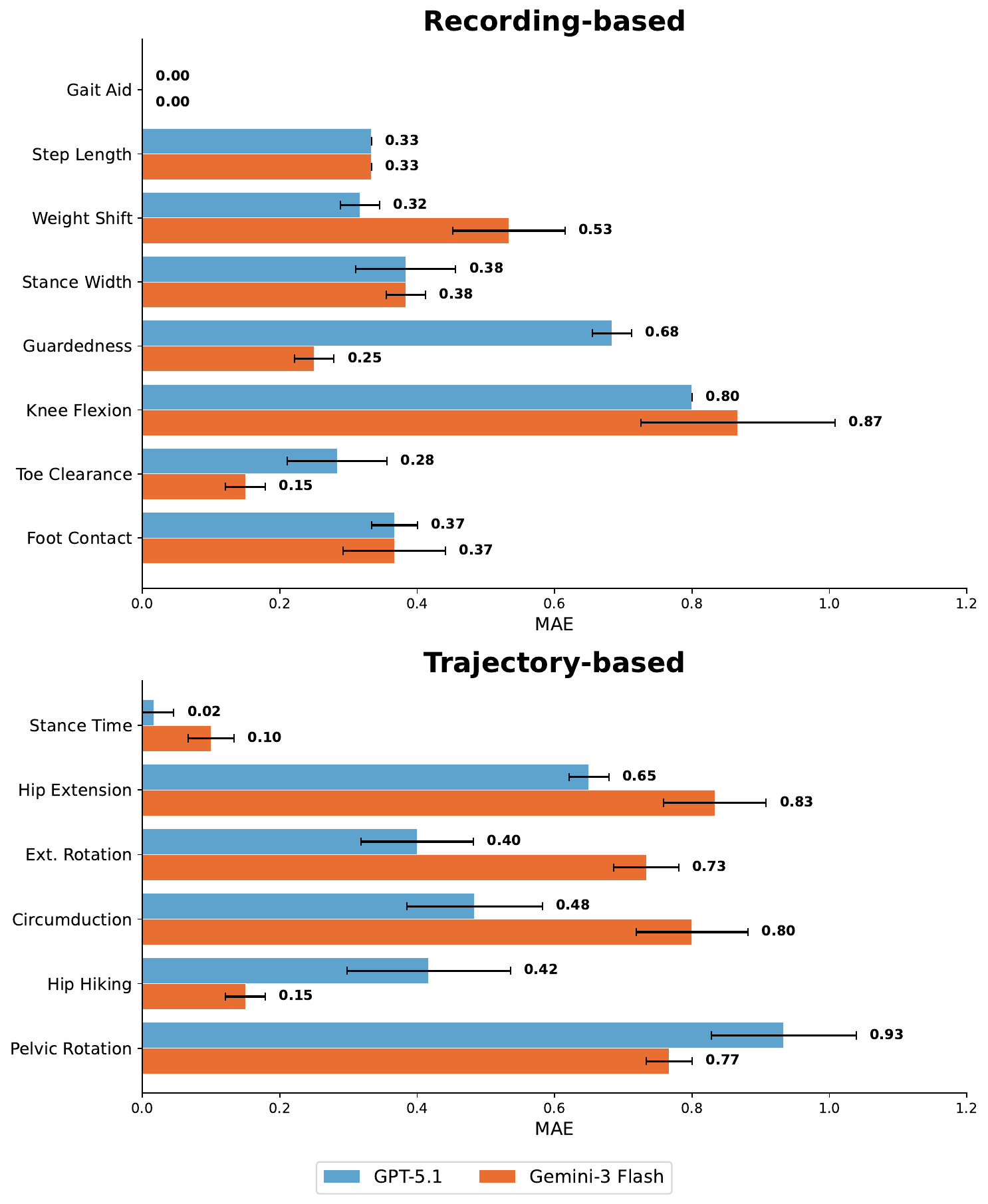}}
  \vspace{-5pt}
  \caption{Per-factor MAE of WGS assessments by GPT-5.1 and Gemini-3 Flash across 45 samples. Each bar represents the MAE between predicted and reference ordinal scores for a given WGS factor. Error bars indicate standard deviation across runs.}
  \label{fig:per_attribute_mae_combined}
  \vspace{-15pt}
\end{figure}

Regarding systematic behavior, the \textbf{predominantly positive Bias values among all backbones indicate a consistent tendency to overestimate WGS scores}, which may reflect the potential miscalibration issues of modern LLMs. From this observation, we highlight the importance of careful evaluation when deploying LLM-based systems in medical applications to mitigate systematic prediction errors.

In Figure~\ref{fig:per_attribute_mae_combined}, we examined the factor-wise MAE between the OGA-AID reports and the reference. While Gemini-3 Flash achieves the lower total MAE, GPT-5.1 is generally more consistent with the reference WGS. Nonetheless, GPT-5.1 can also misalign significantly compared to the reference in some factors, such as guardedness.

In Table~\ref{tab:feature_importances}, we further examined the contributions of different input modalities within OGA-AID, featuring our best LLM performers GPT-5.1 and Gemini-3 Flash. We observe that removing any input component consistently increases MAE, confirming the importance of multimodal integration for accurate WGS estimation. The complete configuration achieves the strongest performance for both GPT-5.1 (MAE = 2.05) and Gemini-3 Flash (MAE = 1.94), whereas relying on Recordings ($R$) only without MoCap trajectories ($T$) leads to a clear performance decline. Notably, under single-modality settings, GPT-5.1 outperforms Gemini-3 Flash, suggesting stronger standalone video and image understanding. In contrast, Gemini-3 Flash demonstrates better interpretation of multiple input modalities, resulting in better performance in the full multimodal setting. However, these results may be subject to the specific prompting strategies employed.

\begin{table}[htbp]
\centering
\caption{Performance under different input combinations for OGA-AID. $R$ denotes movement recordings, $T$ denotes trajectory, $D$ denotes patient clinical profile, \textcolor{BurntOrange}{orange} and \textcolor{NavyBlue}{blue} colors denote negative and positive Bias respectively.}
\vspace{-5pt}
\label{tab:feature_importances}
\setlength{\tabcolsep}{1pt}
\begin{tabular}{lcccccc}
\toprule
\multirow{2}{*}{Inputs} 
& \multicolumn{3}{c}{GPT-5.1} 
& \multicolumn{3}{c}{Gemini-3 Flash} \\
\cmidrule(lr){2-4} \cmidrule(lr){5-7}
& MAE $\downarrow$ & Max AE $\downarrow$ & Bias
& MAE $\downarrow$ & Max AE $\downarrow$ & Bias \\
\midrule

$R$           & 3.80 & 9.50 & \textcolor{BurntOrange}{-3.34}  & 5.44 & 10.75 & \textcolor{BurntOrange}{-5.32} \\
$R$+$D$       & 2.33 & 6.75 & \textcolor{BurntOrange}{-1.53}  & 6.24 & 12.75 & \textcolor{BurntOrange}{-6.15} \\
$R$+$T$       & 2.85 & 8.50 & \textcolor{BurntOrange}{-1.27}  & 2.06 & 5.50 & \textcolor{NavyBlue}{0.48} \\
$R$+$T$+$D$   & \textbf{2.05} & \textbf{6.75} & \textcolor{NavyBlue}{0.38} 
               & \textbf{1.94} & \textbf{4.25} & \textcolor{NavyBlue}{0.23} \\
\bottomrule
\end{tabular}
\vspace{-5pt}
\end{table}

When $R$ is provided without $T$, both GPT-5.1 and Gemini-3 Flash show significant negative biases (-3.34 and -5.32). This suggests that by \textbf{only looking at the visual movement, the LLMs underestimate the stroke severity}. Incorporating MoCap trajectories helps reduce this effect. We also emphasize the importance of the patient’s clinical profile $D$, which enables OGA-AID to identify the hemiparetic side and provides crucial information that guides the Recording Observer in interpreting the patient’s gait.

\subsection{Clinician-in-the-loop Evaluation}
\label{ssec: cil_eval}

Table~\ref{tab:cil_exp1} presents scenarios where clinicians’ preliminary observation notes are provided to OGA-AID as additional context. The results indicate that OGA-AID can leverage expert insights and subsequently improve the quality of the generated reports, as reflected in the overall reduction in MAE for both GPT-5.1 and Gemini-3 Flash compared to the base cases with no initial observations. 

\begin{table}[htbp]
\centering
\caption{Performance with clinician preliminary observations. “Obs” indicates the length of observations supplied to OGA-AID, with "-" referring to the base case without any initial observation.}
\vspace{-5pt}
\label{tab:cil_exp1}
\setlength{\tabcolsep}{1pt}
\begin{tabular}{lcccccc}
\toprule
\multirow{2}{*}{\makecell{Obs}} 
& \multicolumn{3}{c}{GPT-5.1} 
& \multicolumn{3}{c}{Gemini-3 Flash} \\
\cmidrule(lr){2-4} \cmidrule(lr){5-7}
& MAE $\downarrow$ & Max AE $\downarrow$ & Bias
& MAE $\downarrow$ & Max AE $\downarrow$ & Bias \\
\midrule

- & 2.05 & 6.75 & \textcolor{NavyBlue}{0.38} & 1.94 & 4.25 & \textcolor{NavyBlue}{0.23} \\
Short & 2.07 & 6.00 & \textcolor{NavyBlue}{1.01} & 1.90 & 3.25 & \textcolor{BurntOrange}{-0.03}\\
Med. & 1.96 & 8.00 & \textcolor{NavyBlue}{0.89} & 1.76 & 6.00 & \textcolor{NavyBlue}{0.38}\\
Long & 1.67 & 4.00 & \textcolor{NavyBlue}{1.00} & 1.58 & 4.00 & \textcolor{BurntOrange}{-0.20}\\

\bottomrule
\end{tabular}
\vspace{-5pt}
\end{table}

Furthermore, performance improves as the richness of the clinician's initial observations increases. When longer notes are provided, OGA-AID achieves up to 22.7\% reduction in MAE (1.58 compared to 1.94). Medium observations also yield consistent performance gains, though to a lesser extent. In some cases, however, we observe a higher Max AE, suggesting that for certain outlier samples, additional clinical context may introduce ambiguity that affects how the LLM interprets gait characteristics that differ from what the clinician expected. 

Overall, these findings suggest that clinician-provided contextual cues help guide the analysis process of the agentic pipeline. OGA-AID appears to benefit from lightweight expert input that supports interpretation and reduces uncertainty during report generation. This observation highlights the complementary relationship between human clinical judgment and AI-assisted analysis, where brief clinician insights can meaningfully enhance automated assessment.

\section{Discussions}
\label{sec: discussions}

In this section, we reflect on the key findings of our study and discuss several limitations and future directions.

\textbf{Clinical Implications.} Although the proposed system demonstrates encouraging results, substantial opportunities remain to improve performance. Future efforts may focus on more effective prompt engineering strategies and enhanced processing methods tailored to different input modalities, particularly video recordings and MoCap trajectory data. For instance, additional trajectory processing could be applied to provide richer analytical contexts for LLM inferences. Nevertheless, the main contribution of this work is establishing the feasibility of LLM-based, agentic AI in physiotherapy, particularly in OGA and post-stroke rehabilitation. Within this framework, we investigate how LLM-based systems operate in realistic clinical workflows, assessing both analytical performance and the alignment between generated reports and clinician judgment.

\textbf{Data Constraint.} Rehabilitation research is inherently constrained by limited shareable data availability. Unlike many medical domains where large-scale textual or imaging datasets are available, collecting post-stroke motion data is time-consuming due to patients' limited mobility and often requires strict privacy considerations. Additionally, although some gait datasets exist \cite{Liang2020AsianCentricMovement, van_criekinge_full-body_motion_capture_2023}, they typically consist of raw kinematic recordings accompanied by minimal metadata, such as demographic or diagnostic information, and rarely include high-level textual descriptions or clinically meaningful gait assessments. The lack of expert-annotated narrative evaluations poses a significant challenge for the development and validation of LLM-driven gait analysis systems. To mitigate data scarcity and ensure statistical robustness, we upsampled the dataset by collecting multiple trials from the stroke participants. Throughout the evaluation of OGA-AID, we collaborated closely with clinicians to ensure reliable assessment while maintaining alignment with real clinical workflows, avoiding additional workload or disruption to existing practice.

\section{Conclusion}
\label{sec: conclusion}

We presented OGA-AID, a clinician-in-the-loop, multi-agent AI system for post-stroke OGA reports. By integrating gait recordings, MoCap joint trajectories, and clinical context within a structured agentic pipeline, the system generates clinically aligned draft reports while preserving clinician oversight. This work demonstrates the feasibility of agentic LLM systems for supporting clinical assessment workflows, physiotherapy, and rehabilitation practice.

\section*{Acknowledgments}

Ananda Sidarta, Khoi Nguyen, and Koh Hui Yu are supported by tripartite funding from A*STAR/NTU/NHG, within the Rehabilitation Research Institute of Singapore. Nghia D. Nguyen is supported by the Vingroup Science and Technology Scholarship Program.
{
    \small
    \bibliographystyle{ieeenat_fullname}
    \bibliography{main}
}

\end{document}